\documentclass[preprint,superscriptaddress,showpacs]{revtex4}
\usepackage{rotating}
\usepackage[english]{babel}

\newcommand{\EVRY}{D\'epartement de Physique et Mod\'elisation,
Universit\'e d'Evry Val d'Essonne\\
Boulevard F. Mitterrand, 91025 Evry cedex}
\newcommand{\LKB}{Laboratoire Kastler Brossel, Universit\'e Pierre et Marie Curie\\
Case 74, 4 place Jussieu, 75252 Paris, France}

\begin{document}
\title{Quantum three body Coulomb problem in two dimensions}

\author{L. Hilico}
\email{hilico@spectro.jussieu.fr}
\affiliation{\LKB}
\affiliation{\EVRY}

\author{B. Gr\'emaud}
\affiliation{\LKB}

\author{T. Jonckheere}
\affiliation{\LKB}
\affiliation{\EVRY}

\author{N. Billy}
\affiliation{\LKB}
\affiliation{\EVRY}

\author{D. Delande}
\affiliation{\LKB}

\date{\today}

\begin{abstract}
We study the three-body Coulomb problem in two dimensions and
show how to calculate very accurately its quantum properties. The use
of a convenient set of coordinates makes it possible to write 
the Schr\"{o}dinger 
equation only using annihilation and creation operators
of four harmonic oscillators, coupled by various terms of degree
up to twelve.
We analyse in details the discrete symmetry properties of the
eigenstates. 
The energy levels and eigenstates of the two-dimensional 
helium atom are obtained
numerically, by expanding
the Schr\"{o}dinger equation  on a convenient basis set,
that gives sparse banded matrices, 
and thus opens up the way to accurate and efficient calculations. We give some 
very accurate values of the energy levels of the first bound Rydberg series.
Using the complex coordinate method, we are also able to calculate energies and widths
of doubly excited states, i.e. resonances above the first ionization threshold.
For the two-dimensional $H^{-}$ ion, only one bound state is found.
\end{abstract}


\maketitle
\section{Introduction}

Since the very beginning of quantum mechanics, the helium atom has attracted
much attention as it is one of the simplest system where the Schr\"odinger equation
cannot be solved exactly. Recently, it has been understood that the lack of 
an exact solution is the direct quantum counterpart of the non-integrable
character of the corresponding classical dynamics~\cite{Rost}. Indeed, it has been
discovered that, for most initial conditions (positions and velocities of the
two electrons), the classical dynamics is chaotic, with the total energy and
the total angular momentum being the only constants of motion. Together with the
development of sophisticated numerical methods for computing the quantum energy levels
\cite{gremaud,burgers,goldman,drake,korobov,chuluunbaatar}, there have been major improvements on semiclassical techniques
which allow to compute approximate values of the energy levels from the
knowledge of the classical dynamics. The most dramatic success is the
use of periodic orbit theory, where the energy levels are calculated from
simple properties (action, period, stability...) of a 
(preferably large) set of classical periodic orbits \cite{Rost}. Most of the
quantum and semiclassical calculations concentrated on states with low
total angular momentum for at least two reasons: firstly, these are the states
experimentally prepared when using an optical excitation from a low excited state
and, secondly, this is the situation where the classical dynamics is well
known. 

Of special interest are the S states with zero total angular momentum.
Classically, the motion of the two electrons takes place in a fixed plane.
Thus, the classical dynamics is fully identical with the classical
dynamics of the two-dimensional (2D) helium atom. It turns out that,
although it seems to be a simpler system, there has been only very little
interest in this 2D three body Coulomb problem and practically
no quantum calculation. It is the aim of this paper to fill this hole.
It can be also expected that, when a ``real" 3D helium atom with 
low (or zero) initial momentum is exposed
to an external perturbation, its response will not be very different
from the one of the 2D atom, provided angular momentum does not play
a crucial role in the physical processes involved.
For example, when a helium atom is exposed to a strong non-resonant
low-frequency electromagnetic field, it may absorb a large number of photons
leading eventually to single or even double ionization. It seems likely
that the correlation between the two electrons plays a major role
in this process (especially in the generation of high harmonics
of the electromagnetic field), while the total angular momentum remains
relatively small. Another example is the production of doubly ionized atoms
where a process involving symmetric excitation of the two electrons (with
zero total angular momentum) has been recently proposed \cite{sacha}.
In these situations, the full 3D quantum calculation for such a system is
not presently feasible, except for the very lowest states. On the other hand,
a 2D quantum calculation seems reachable. This would allow to determine whether the
proposed process is relevant or not. It is thus highly desirable to
be able to compute accurately the quantum properties of the 2D helium atom.
 
A second motivation to study the 2D three body Coulomb problem comes from 
semi-conductor physics. The study of excitons - the bound aggregate of 
an electron from the conduction band 
and a hole from the valence band, each particle with a given effective mass-
 is an important tool to study semi-conductors. In 1958, M.A. Lampert \cite{lampert}  
 has shown that 
three body complexes called trions (an electron or a hole bound to an exciton) should be 
 observable at low temperatures, and this was confirmed later by variational calculations, 
showing the stability of trions against dissociation into a exciton 
and a free electron or a hole 
(see \cite{stebe} for references). Since, the progress in semiconductor 
technology have made possible 
the fabrication of quasi 2D systems. It was then realized \cite{Bobrysheva,stebe} 
that in such systems,
 trions would  have an increased stability due to the 2D confinement, 
 and should thus be more easily 
observable. The trions are responsible for satellites on the excitonic lines in 
luminescence spectra. Several observations have been reported since 
the first one in 1993 
\cite{kheng,Finkelstein,buhmann,shields,paganotto,huard}, and 
compared with theoretical predictions \cite{stebe,whittaker}. 
In this context, a precise calculation 
of the energy levels of the excitonic trions in a 2D system as a 
function of the ratio of the 
effective masses, with and without external field, is highly 
valuable, and justifies the 
 methods and  calculations introduced in this paper. 
The 2D hydrogen molecular ion $H_2^+$ has also 
been studied in the frame 
of the Born Oppenheimer approximation in Ref. \cite{zhu}, where the first two 
electronic energy curves are given.  

The paper is organized as follows. In section \ref{Schro-eq}, 
we discuss the physical symmetries of the 
2D three body Coulomb problem. We then introduce a new set of 
parabolic-like coordinates, give the expression of the Hamiltonian operator
and show that we can find a basis in which the Schr\"{o}dinger equation 
involves sparse banded matrices, allowing accurate numerical calculations.
In section \ref{Disc-sym}, we analyse the group structure of the discrete symmetries 
of the new Hamiltonian, showing that the complications introduced
by the not one-to-one character of the 
change of coordinates can be taken into account exactly
and actually does not lead to any difficulty.
In section \ref{Num-sol}, we first explain the detailed structure 
of the basis set that we use. We then discuss the structure 
of the expected energy spectrum 
in the case of a 2D helium atom with an infinite mass nucleus,
and give the energies of the lowest levels 
in the bound Rydberg series, as well as -- using the technique
of complex coordinates -- the energy and width of the 
first doubly excited resonance.

\section{The Schr\"{o}dinger equation\label{Schro-eq}}
\subsection{Hamiltonian}
The three body problem in two dimensions has 6 degrees of 
freedom that can be reduced to 4 in the center of mass frame.
Here, as depicted in figure \ref{coord}, ${\bf r}_1$ and ${\bf r}_2$ denote the 
positions of particle 1 
or 2 with respect to particle 3, and ${\bf p}_1$ and ${\bf p}_2$ 
the conjugate momenta. In atomic units (such
that $\hbar$, $4\pi\epsilon_0$, the mass $m$ of the electron and the
elementary charge are all equal to unity), the Hamiltonian writes,
neglecting QED and relativistic effets:

\begin{figure}
\includegraphics[width=9cm]{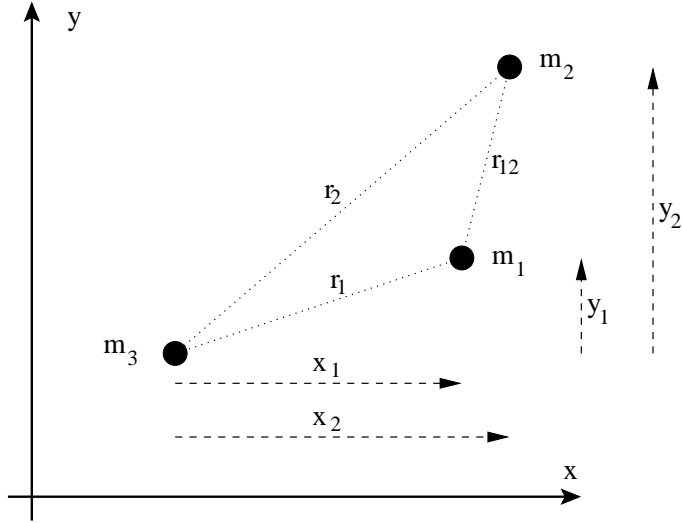}
\caption{\label{coord} The relative cartesian
 coordinates of particles 1 and 2 with respect to particle 3 
 are $(x_1,y_1)$ and $(x_2,y_2)$.
The interparticles distances $r_1$, $r_2$ and $r_{12}$.}
\end{figure}

\begin{equation}
H=\frac{{\bf p}_1^{\ 2}}{2\mu_{13}}+\frac{{\bf p}_2^{\ 2}}{2\mu_{23}}
+\frac{{\bf p}_1.{\bf p}_2}{m_3}
+\frac{Q_1Q_3}{r_1}+\frac{Q_2Q_3}{r_2}+\frac{Q_1Q_2}{r_{12}},
\label{hamil1}
\end{equation} 
where $m_3$ is the mass of the third particle 
(in unit of the electron mass), and 
$\mu_{13}$ (resp. $\mu_{23}$) is the reduced mass of 
particle 1 (resp. 2) and particle 3. 
$Q_1$, $Q_2$ and $Q_3$ are the charges of the particules in unit of the elementary charge.
$r_{12}$ 
is the distance between the particles 1 and 2.
The 2D helium atom with a fixed nucleus corresponds to the case 
where $m_3$ is infinite, $\mu_{13}=\mu_{23}=1$, $Q_1=Q_2=-1$ and $Q_3=2$. 

As for the 3D three body problem \cite{richter}, 
we regularize the Schr\"{o}dinger 
equation, i.e. remove the denominators, by multiplying it by $16\ r_1r_2r_{12}$.
The eigenstate $|\Psi\rangle$ with energy $E$ then satisfies
the generalized linear eigenequation:
\begin{equation}
16 r_1r_2r_{12}\left(\frac{{\bf p}_1^{\ 2}}{2\mu_{13}}+\frac{{\bf p}_2^{\ 2}}{2\mu_{23}}
+\frac{{\bf p}_1.{\bf p}_2}{m_3}\right)
|\Psi\rangle +V|\Psi\rangle
=16\ r_1r_2r_{12}\ E\ |\Psi\rangle, \label{linear-pb}
\end{equation}
where 
\begin{equation}
V=16\ (Q_1Q_3\ r_2r_{12}+Q_2Q_3\ r_1r_{12}+Q_1Q_2\ r_1r_2).\label{potentiel}
\end{equation}

\subsection{Symmetries \label{sym-phys}}
The symmetries of the 2D three body problem are the rotational 
invariance around an axis $(\Delta)$ perpendicular to the plane, the parity $\Pi$ and, 
when particles 1 and 2 are identical, the exchange symmetry $P_{12}$.
In two dimensions, the parity operator $\Pi$ coincides with a rotation of angle $\pi$ 
around $(\Delta)$, 
so that $\Pi$ and the angular momentum $L_z$ are related by
\begin{equation}
\Pi=(-1)^{L_z}.
\end{equation}
We also introduce the two commuting symmetries 
$\Pi_x$ (symmetry with respect to the $x$ axis)
and $\Pi_y$ (symmetry with respect to the $y$ axis).
They are related to total parity through $\Pi_x \Pi_y=\Pi_y \Pi_x=\Pi$.
The group generated by $\Pi_x$, $\Pi_y$ and $P_{12}$ is the so called $D_{2h}$ 
point group. It is an invariance group of the Hamiltonian (\ref{hamil1}),
for identical particles 1 and 2. 
The symmetries $\Pi_x$ and $\Pi_y$ both commute with parity, 
but not with the angular momentum since, for instance, 
$\Pi_x\ L_z=-L_z \Pi_x$.
As a consequence, the eigenstates of the 2D three body Coulomb problem 
can be labelled by their angular momentum 
$M_L=0,\ \pm1,\ \pm2\ ...$ and by the exchange symmetry when 
particles 1 and 2 are identical. 
The spectrum corresponding to $M_L$ and $-M_L$ angular momenta 
are identical: this (Kramers) degeneracy is a direct consequence of the time reversal 
invariance of the problem \cite{landau}.
Alternatively, the eigenstates could also be labelled by parity 
with respect 
to the $x$ axis and the absolute value of the angular momentum.

When the system is exposed to an external uniform electric field along 
the $x$ axis,
the angular momentum is no longer preserved. The only remaining symmetries 
are $\Pi_x$ and $P_{12}$.


\subsection{Parabolic coordinates}
In order to perform efficient and accurate numerical calculations, we wish to 
obtain a sparse banded matrix representation of 
the linear problem (\ref{linear-pb})
where the non-zero matrix elements are known in a closed form. 
We thus have to find a basis set 
in which the various terms of the Hamiltonian have strong selection rules. This can be 
achieved for example if all terms of the Hamiltonian can be expanded in 
polynomial combinations of position and (conjugate) momentum coordinates: in such
a case, the set of eigenstates of an harmonic oscillator is convenient. Our situation
is slightly more complicated, because the Hamiltonian involves the
interparticle distance. How to deal with such a problem is well known
for the hydrogen atom: by introducing a set of so-called parabolic or
semiparabolic coordinates \cite{englefield}, one can map the 2D hydrogen atom
on an harmonic oscillator. The method used here for the 2D helium atom
is inspired by such a treatment, although it is technically more
complicated.

If $x$ and $y$ are the cartesian coordinates of a point in a 2D space 
and $z=x+iy$ is the associated complex number, the distance from the origin 
is  $r=|z|=\sqrt{x^2+y^2}$, and its expression involves 
a square root function. 
The square root can be removed if we introduce the complex variable $Z=X+i Y$ 
defined by $z=\frac{Z^2}{2}$, since $r=\frac{|Z|^2}{2}=\frac{X^2+Y^2}{2}$. 
$X$ and $Y$ are the 
parabolic coordinates, related to $x$ and $y$ by:
\begin{equation}
x=\frac{X^2-Y^2}{2} \ \mathrm{and} \ y=XY.\label{XYxy}
\end{equation}
The parabolic coordinates are extremely convenient to represent 
the hydrogen atom in two dimensions \cite{englefield}, or the Stark effet 
of the 3D hydrogen atom \cite{landau}. Of course, the correspondance 
between $(X,Y)$ and $(x,y)$ given in equation (\ref{XYxy}) is not one to one.
The difficulties related to that choice 
of coordinates are discussed in section \ref{disc-parabolic}.

We now come to the case of three particles. The complex positions of 
particles 1 and 2 with respect to particle 3 are $z_1$ and $z_2$, 
and $Z_1$ and $Z_2$ are the associated parabolic coordinates. 
The interparticle distances then write $r_1=\frac{|Z_1|^2}{2}$, 
$r_2=\frac{|Z_2|^2}{2}$ and then $r_{12}=|z_1-z_2|=\frac{|(Z_1+Z_2)|\ |(Z_1-Z_2)|}{2}$.
If we introduce the two complex numbers 
$Z_p=\frac{Z_1+Z_2}{\sqrt{2}}$ and $Z_m=\frac{Z_1-Z_2}{\sqrt{2}}$, 
the distance $r_{12}$ appears as the product of the moduli of $Z_p$ and $Z_m$.
Since we want to express $r_{12}$ using square moduli, we introduce a second parabolic 
transformation on both $Z_p$ and $Z_m$ by setting 
$Z_p=\frac{\Xi_p^2}{2}$ and $Z_m=\frac{\Xi_m^2}{2}$. The 
three distances 
are then expressed as the square moduli:
\begin{eqnarray}
r_1&=&\frac{1}{16}\left|\Xi_p^2+\Xi_m^2 \right|^2, \\
r_2&=&\frac{1}{16}\left|\Xi_p^2-\Xi_m^2 \right|^2, \\
r_{12}&=&\frac{1}{4}\left|\Xi_p \Xi_m \right|^2. 
\end{eqnarray}
As a consequence, the three distances have polynomial expressions when 
 they are expressed 
with the new coordinates $(x_p,y_p,x_m,y_m)$ defined by 
$\Xi_p=x_p+iy_p$ and $\Xi_m=x_m+iy_m$. Those coordinates are related 
to the 
initial cartesian coordinates $(x_1,y_1,x_2,y_2)$ by :
\begin{eqnarray}
x_1&=&\frac{1}{16}\left(x_p^2-y_p^2-2x_py_p +x_m^2-y_m^2-2x_my_m\right)
\left(x_p^2-y_p^2+2x_py_p +x_m^2-y_m^2+2x_my_m \right),  \nonumber\\
y_1&=&\frac{1}{4}\left(x_p^2-y_p^2 +x_m^2-y_m^2\right)  
\left(x_py_p +x_my_m \right),  \nonumber\\
x_2&=&\frac{1}{16}\left(x_p^2-y_p^2+2x_py_p -x_m^2+y_m^2-2x_my_m\right)
\left(x_p^2-y_p^2-2x_py_p -x_m^2+y_m^2+2x_my_m \right),  \nonumber\\
y_2&=&\frac{1}{4}\left(x_p^2-y_p^2 -x_m^2+y_m^2\right)  
\left(x_py_p -x_my_m \right),  \label{transf}
\end{eqnarray}
and the three distances are:
\begin{eqnarray}
r_1&=&\frac{1}{16}\left((x_p-y_m)^2+(y_p+x_m)^2\right)
\left((x_p+y_m)^2+(y_p-x_m)^2\right), \nonumber\\
r_2&=&\frac{1}{16}\left((x_p+x_m)^2+(y_p+y_m)^2\right)
\left((x_p-x_m)^2+(y_p-y_m)^2\right), \nonumber\\
r_{12}&=&\frac{1}{4}\left(x_p^2+y_p^2\right)\left(x_m^2+y_m^2\right).\label{distances}
\end{eqnarray}

\subsection{The Schr\"odinger equation}
The Schr\"odinger equation (\ref{linear-pb}) can be written as:
\begin{equation}
\left\{\frac{T_1}{2\mu_{13}}+\frac{T_2}{2\mu_{23}}
+\frac{T_{12}}{m_3}
+V\right\}|\Psi(x_p,y_p,x_m,y_m)\rangle=\ E\ B\ |\Psi(x_p,y_p,x_m,y_m)\rangle, \label{Schro-equ}
\end{equation}
where the kinetic energy terms are:
\begin{eqnarray}
T_1&=&-\frac{1}{16}\left((x_p+x_m)^2+(y_p+y_m)^2\right)\left((x_p-x_m)^2+(y_p-y_m)^2\right) \nonumber\\
&&\,\,\left\{
(x_m^2+y_m^2)\left(\frac{{\partial}^2}{\partial x_p^2}+\frac{{\partial}^2}{\partial y_p^2}\right)
+(x_p^2+y_p^2)\left(\frac{{\partial}^2}{\partial x_m^2}+\frac{{\partial}^2}{\partial y_m^2}\right) \right. \nonumber\\
&&\ \left.+2(x_px_m+y_py_m)\left(\frac{{\partial}^2}{\partial x_p\partial x_m}+\frac{{\partial}^2}{\partial y_p\partial y_m}\right)
-2(x_py_m-y_px_m)\left(\frac{{\partial}^2}{\partial x_p\partial y_m}-\frac{{\partial}^2}{\partial y_p\partial x_m}\right)  
\right\},  \nonumber\\
T_2&=&-\frac{1}{16}\left((x_p-y_m)^2+(y_p+x_m)^2\right)\left((x_p+y_m)^2+(y_p-x_m)^2\right) \nonumber\\
&&\,\,\left\{
(x_m^2+y_m^2)\left(\frac{{\partial}^2}{\partial x_p^2}+\frac{{\partial}^2}{\partial y_p^2}\right)
+(x_p^2+y_p^2)\left(\frac{{\partial}^2}{\partial x_m^2}+\frac{{\partial}^2}{\partial y_m^2}\right) \right. \nonumber\\
&&\ \left.-2(x_px_m+y_py_m)\left(\frac{{\partial}^2}{\partial x_p\partial x_m}+\frac{{\partial}^2}{\partial y_p\partial y_m}\right)
+2(x_py_m-y_px_m)\left(\frac{{\partial}^2}{\partial x_p\partial y_m}-\frac{{\partial}^2}{\partial y_p\partial x_m}\right)  
\right\},  \nonumber\\
T_{12}&=&-\frac{1}{16}\left((x_p^2+y_p^2)^2-(x_m^2+y_m^2)^2\right)
\left\{(x_m^2+y_m^2)\left(\frac{{\partial}^2}{\partial x_p^2}+\frac{{\partial}^2}{\partial y_p^2}\right)
-(x_p^2+y_p^2)\left(\frac{{\partial}^2}{\partial x_m^2}+\frac{{\partial}^2}{\partial y_m^2}\right) \right\} \nonumber\\
&&-\frac{1}{2}(x_px_m+y_py_m)(x_py_m-y_px_m)
\left\{(x_px_m+y_py_m)\left(\frac{{\partial}^2}{\partial y_p\partial x_m}-\frac{{\partial}^2}{\partial x_p\partial y_m}\right) \right. \nonumber\\
&&\left.-(x_py_m-y_px_m)\left(\frac{{\partial}^2}{\partial x_p\partial x_m}+\frac{{\partial}^2}{\partial y_p\partial y_m}\right) \right\},\nonumber\\
B&=&16\ r_1r_2r_{12}.\label{Bmat}
\end{eqnarray}
The expressions of $B$ and $V$ 
can be deduced from equations (\ref{potentiel}) and (\ref{distances}). 
The Jacobian of the coordinate transformation is $16\ r_1r_2r_{12}$. 
The scalar product of two wave functions is given in appendix \ref{appendix-pente}.

The various terms in the Schr\"{o}dinger equation (\ref{Schro-equ})
are polynomials in the coordinates $(x_p,y_p,x_m,y_m)$ and their
associated momenta (partial derivatives $-i\partial\!/\!\partial \{x_p,y_p,x_m,y_m\}$). 
The operators 
$T_1$, $T_2$, $T_{12}$, $V$ and $B$ 
can thus be expressed using the corresponding annihiliation and creation operators:
\begin{equation}
a_{x_p}=\frac{1}{\sqrt{2}}\left(x_p+\frac{\partial}{\partial x_p}\right)\ ,
\  a_{x_p}^{\dagger}=\frac{1}{\sqrt{2}}\left(x_p-\frac{\partial}{\partial x_p}\right). \label{axp}
\end{equation}
This shows that the 2D three body Coulomb problem can be described 
using the annihilation and creation operators of 4 harmonic oscillators. 
The Hamiltonian is a 
polynomial of  degree 12 in the annihilation and creation operators.
Consequently, it will be possible to choose a basis of tensorial products of 
Fock states of each harmonic oscillator, for which the operators involved 
in the Schr\"{o}dinger equation exhibit strong coupling rules.

From the annihilation and creation operators associated with the new coordinates,
we introduce the right and left circular operators in the planes
 $(x_p,y_p)$ and $(x_m,y_m)$ defined by:
\begin{eqnarray}
 a_1&=&(a_{x_p}-ia_{y_p})/\sqrt{2}, \nonumber\\
 a_2&=&(a_{x_p}+ia_{y_p})/\sqrt{2}, \nonumber\\
 a_3&=&(a_{x_m}-ia_{y_m})/\sqrt{2}, \label{circular-op}\\
 a_4&=&(a_{x_m}+ia_{y_m})/\sqrt{2}. \nonumber
\end{eqnarray}
Using the symbolic calculation 
language {\it Maple V}, we have calculated the normal ordered 
expression of the various 
operators involved in the Hamiltonian. Those expressions are too 
long to be published here. Indeed, the operators $T_1$ and $T_2$ contain 625 terms, 
$T_{12}$ 331, the potential operators $r_1r_{12}$ and $r_2r_{12}$ 517, 
$r_1r_2$ 159 and $B$
 1463. When particles 1 and 2 are identical, 
 the Hamiltonian involves the kinetic term $T_1+T_2$ and the potential term
 $(r_1+r_2)r_{12}$ that have only 335 and 275 terms, because the terms
 of $T_1$ and $T_2$ that do not commute with the exchange operator $P_{12}$
 cancel out. 

\subsection{Angular momentum}
The angular momentum $L_z$ has a very simple expression when expressed with the
$(x_p,y_p,x_m,y_m)$ coordinates:
\begin{eqnarray}
L_z&=&-i\left(x_1\frac{\partial}{\partial y_1}-y_1\frac{\partial}{\partial x_1}
+x_2\frac{\partial}{\partial y_2}-y_2\frac{\partial}{\partial x_2}\right) \nonumber\\
L_z&=&-\frac{i}{4}\left(x_p\frac{\partial}{\partial y_p}-y_p\frac{\partial}{\partial x_p}
+x_m\frac{\partial}{\partial y_m}-y_m\frac{\partial}{\partial x_m}\right).\label{lz}
\end{eqnarray}
The relation $z=\frac{Z^2}{2}$ between the cartesian and the parabolic complex 
numbers shows that a rotation of angle $\theta$ on $Z$ is a rotation
of $2\theta$ on $z$. Consequently, a factor of $2$ appears in the expression 
of the angular momentum in parabolic coordinates \cite{englefield}. 
Since we have performed two successive parabolic transformations
to define the $(x_p,y_p,x_m,y_m)$ coordinates, we have a factor 4 in the denominator 
of equation (\ref{lz}). 
 With the annihilation and creation operators (\ref{circular-op}), 
 the angular momentum simply writes :
\begin{equation} 
L_z=(N_1-N_2+N_3-N_4)/4,
\end{equation}
where
the number operators are $N_i=a_i^{\dagger}a_i$. They are related to the number
operators corresponding the the annihilation and creation
operators $a_{x_p}$, ... given in eq.~(\ref{axp}) by:
\begin{eqnarray}
N_1+N_2&=&N_{x_p}+N_{y_p},\nonumber\\
N_3+N_4&=&N_{x_m}+N_{y_m}. \label{relation-N}
\end{eqnarray}
\section{Discrete symmetries\label{Disc-sym}}
\subsection{Physical symmetries}
The Hamiltonian (\ref{hamil1}) has two discrete symmetries, $\Pi_x$ and $\Pi_y$,
which are the symmetries with respect to two orthogonal axis 
in the physical plane. Using the new $(x_p,y_p,x_m,y_m)$ coordinates,
 they can be expressed for instance as:
\begin{equation}
\begin{array}{cc}
\begin{array}{llcc}
\Pi_x : &x_p &\rightarrow &x_p\\
         &y_p &\rightarrow &-y_p\\
	 &x_m &\rightarrow &x_m\\
	 &y_m &\rightarrow &-y_m
\end{array}
&
\begin{array}{llcc}
\Pi_y : &x_p &\rightarrow &(x_p+y_p)/\sqrt{2}\\
         &y_p &\rightarrow &(x_p-y_p)/\sqrt{2}\\
	 &x_m &\rightarrow &(x_m+y_m)/\sqrt{2}\\
	 &y_m &\rightarrow &(x_m-y_m)/\sqrt{2}.
\end{array}
\end{array}\label{pixpiy}
\end{equation}
Moreover, if particles 1 and 2 are identical, the Hamiltonian commutes 
with the 
exchange operator $P_{12}$.
The effect of $P_{12}$ on the $(x_p,y_p,x_m,y_m)$ coordinates is:
\begin{equation}
\begin{array}{llcc}
P_{12} : &x_p &\rightarrow &x_p\\
         &y_p &\rightarrow &y_p\\
	 &x_m &\rightarrow &y_m\\
	 &y_m &\rightarrow &-x_m.\label{P12}
\end{array}
\end{equation}
Obviously, the Schr\"{o}dinger equation (\ref{Schro-equ}) written with the $(x_p,y_p,x_m,y_m)$
coordinates is invariant under these transformations.

\subsection{\label{disc-parabolic}``Additionnal" symmetries}
In this section, we analyse the constraints that the physical wave 
functions must satisfy. We first recall what happens 
in the case of a single parabolic transformation.
The parabolic transformation $(X,Y)\rightarrow(x,y)$ defined in equation
(\ref{XYxy}) is a one-to-one mapping of the quarter of 
plane $(X \geq 0, Y \geq 0)$ onto the 
half-plane $(x, y \geq 0)$.
Here, the transformation is used to represent 
 the full cartesian plane $(x,y)$ by extending the domains of $X$ and $Y$ 
 to $]-\infty,+\infty[$. That way, 
we obtain a double mapping of the cartesian plane 
since $(X,Y)$ and $(-X,-Y)$ are mapped on the same point. 
Consequently, the Hamiltonian written with the parabolic coordinates 
has a new discrete symmetry $(X,Y) \rightarrow (-X,-Y)$, 
i.e. the parity with respect to $(X,Y)$.
The physical wave function must be a single-valued function 
of the initial coordinates $(x,y)$ i.e. must fulfill $\Psi(X,Y)=\Psi(-X,-Y)$. 
 Any function of $\Psi(X,Y)$ which satisfies the
Schr\"odinger equation written in the $(X,Y)$ coordinates but does not obey 
the constraint $\Psi(X,Y)=\Psi(-X,-Y)$ is to be rejected as an unphysical solution.

In the particular case where the 
wave function is expanded on a basis built with tensorial products 
of harmonic oscillator eigenstates:
\begin{equation}
|\Psi\rangle=\sum_{n_X,n_Y}{C_{n_X,n_Y}\ |n_X\rangle \otimes |n_Y\rangle},\label{nx+ny}
\end{equation} 
the physical wave function expansion of equation (\ref{nx+ny}) 
is restricted to the even values of $n_X+n_Y$, because the 
parity of the Fock state $|n\rangle$ is $(-1)^n$ \cite {englefield}.

This property can be extended to the case of the transformation 
given in equation (\ref{transf}) that give the cartesian coordinates 
versus the new coordinates $(x_p,y_p,x_m,y_m)$.
Because we perform four parabolic transformations to obtain 
the $(x_p,y_p,x_m,y_m)$ coordinates from the initial cartesian coordinates,
there are four ``additionnal" discrete symmetries which 
leave the Schr\"{o}dinger equation (\ref{Schro-equ}) invariant. 
We denote them $\Pi_1$ defined as $(X_1,Y_1)\rightarrow (-X_1,-Y_1)$, 
$\Pi_2$,
$\Pi_p$ and $\Pi_m$. The effects of those symmetries on the 
$(x_p,y_p,x_m,y_m)$ coordinates are:
\begin{equation}
\begin{array}{cc}
\begin{array}{llcc}
\Pi_1 : &x_p &\rightarrow &-y_m\\
         &y_p &\rightarrow &x_m\\
	 &x_m &\rightarrow &-y_p\\
	 &y_m &\rightarrow &x_p
\end{array}
&
\begin{array}{llcc}
\Pi_2 : &x_p &\rightarrow &x_m\\
         &y_p &\rightarrow &y_m\\
	 &x_m &\rightarrow &x_p\\
	 &y_m &\rightarrow &y_p
\end{array}\\ 
\\
\begin{array}{llcc}
\Pi_p : &x_p &\rightarrow &-x_p\\
         &y_p &\rightarrow &-y_p\\
	 &x_m &\rightarrow &x_m\\
	 &y_m &\rightarrow &y_m
\end{array}
&
\begin{array}{llcc}
\Pi_m : &x_p &\rightarrow &x_p\\
         &y_p &\rightarrow &y_p\\
	 &x_m &\rightarrow &-x_m\\
	 &y_m &\rightarrow &-y_m.
\end{array}
\end{array}
\label{pi1pi2pippim}
\end{equation}
\subsection{Symmetries of the wave function}
The group G generated by the $\Pi_x$, $\Pi_y$, $P_{12}$ and 
the $\Pi_1$, $\Pi_2$, $\Pi_p$, $\Pi_m$ symmetries is an 
invariance group of the Schr\"{o}dinger equation (\ref{Schro-equ}).
It is studied
in details and
its character table is given in appendix \ref{table-xi}.

In order to be singlevalued in the geometrical space $(x_1,y_1,x_2,y_2)$,
the wave function $\Psi(x_p,y_p,x_m,y_m)$ must be invariant 
under any ``additional" symmetry introduced by the non one-to-one change of coordinates, 
i.e. under any of the 
transformations $\Pi_1$, $\Pi_2$, $\Pi_p$, $\Pi_m$.
Then, the wave function must belong to an irreductible representation of G
for which the character of any ``additional" symmetry 
is equal to its dimension. There are only 8 representations with 
this property, all being one-dimensional, that
correspond to the first 8 lines of the character table given in
 appendix (\ref{table-xi}). Consequently, the physical 
 eigenfunctions $\Psi(x_p,y_p,x_m,y_m)$ can be distinguished only by their 
 symmetry properties with respect to $\Pi_x$, $\Pi_y$ and $P_{12}$.
 The 8 physical irreductible representations of G are those of the 
 group $D_{2h}$ (or $mmm$), of order 8, already mentionned 
 in section (\ref{sym-phys}).
 The application 
 that maps each ``additional" symmetry on the identity 
 is a group homomorphic mapping of G on $D_{2h}$.
 
 Finally, we have shown here that all energy levels belong
 to a one-dimensional representation of the discrete symmetry group
 of the Schr\"{o}dinger equation, 
 and are thus expected to be non degenerate (except for the $(M_L,-M_L)$
 mentionned above).
Moreover, using the $(x_p,y_p,x_m,y_m)$ coordinates does not introduce 
extra representations which cannot be distinguished from the physical ones.
The wave functions can be described using a basis exhibiting
the relevant symmetry properties with respect to $\Pi_x$, $\Pi_y$ and
$P_{12}$, or $L_z$ and $P_{12}$. The second feature will be extensively 
used in the numerical implementation. 

In other words, among all solutions of the Schr\"odinger equation (\ref{Schro-equ})
in the $(x_p,y_p,x_m,y_m)$ coordinates, sorting out the unphysical
solutions is rather straightforward and one is left only with the
physical symmetries of the initial system.

\section{Numerical solution\label{Num-sol}}
\subsection{Basis set}
\subsubsection{Basis structure}
To perform numerical calculations of the eigenenergies and 
eigenstates of the three body Coulomb problem, we expand the Schr\"{o}dinger 
equation on a convenient basis, and then solve a linear eigenvalue problem.
Because the different terms of the Hamiltonian have polynomial expressions
in the annihilation and creation operators, we obtain strong 
selection rules
if we choose basis functions that are tensorial products 
of Fock states $|n_i\rangle$ of the harmonic 
oscillator described by the circular
annihilation operator $a_i$. Namely, we set :
\begin{equation}
|n_1,n_2,n_3,n_4\rangle=|n_1\rangle\otimes|n_2\rangle \otimes|n_3\rangle\otimes|n_4\rangle.
\end{equation} 
The indices $n_i$ are then positive integers.
The basis functions are eigenfunctions of the angular momentum, 
corresponding to the integer eigenvalue:
\begin{equation}
M_L=(n_1-n_2+n_3-n_4)/4 \label{ml}.
\end{equation}

The wavefunctions of these basis states are simple. Indeed, they 
are just eigenstates of an harmonic oscillator along the various coordinates.
In the $(x_p,y_p,x_m,y_m)$ coordinates, they should appear as products of
Hermite polynomials and Gaussian functions of the coordinates.
As we use circular creation-annihilation operators, equation~(\ref{circular-op}),
the associated eigenstates of the two-dimensional harmonic oscillators
in the $(x_p,y_p)$ and $(x_m,y_m)$ planes are easily written
in polar coordinates as the product of a $\exp(i\phi)$ term with
an exponential and a Laguerre polynomial of the squared radius. 
The explicit expressions of such states can be found in~\cite{englefield}.

We have previously shown that 
the two successive parabolic transformations introduce 
``additional" unphysical states.
The physical solutions can be 
selected using a basis set that is even with respect to 
all the ``additional" symmetries. 
This choice is performed in two steps.
First, both $n_1+n_2$ and $n_3+n_4$ have to be even numbers. 
Indeed, from equation (\ref{relation-N}), 
$n_1\!+\!n_2=n_{x_p}\!+\!n_{y_p}$ and 
$n_3\!+\!n_4=n_{x_m}\!+\!n_{y_m}$
and the even representations for
$\Pi_p$ and $\Pi_m$ corresponds to even values of 
$n_{x_p}\!+\!n_{y_p}$ and $n_{x_m}\!+\!n_{y_m}$.
Secondly, because the 
transformation $(1,2,3,4)\rightarrow(3,4,1,2)$ 
on the annihiltion and creation operators commutes with the Hamiltonian and 
corresponds to the identity in the physical space, 
the basis functions have to be chosen as 
the symmetric combinations:
\begin{equation}
|n_1,n_2,n_3,n_4\rangle^+=|n_1,n_2,n_3,n_4\rangle+|n_3,n_4,n_1,n_2\rangle.\label{sym-ket}
\end{equation}
Of course, this symmetrised state remains an eigenstate of the angular momentum,
with the same eigenvalue $M_L$.
Taking into account the even parity of $n_1+n_2$ and $n_3+n_4$, 
and thus of $n_1-n_2$ and $n_3-n_4$, and the expression of $M_L$,
we obtain that $n_1-n_2$ (mod 4) 
and $n_3-n_4$ (mod 4) are simultaneously equal to either $0$  or  $2$.
We then set:
\begin{equation}
C_{12}=(n_1-n_2)\ (\mathrm{mod}\ 4)=(n_3-n_4)\ (\mathrm{mod}\ 4).
\end{equation}
When particles 1 and 2 are identical, 
the Hilbert space can be split into a singlet subspace 
corresponding to $C_{12}=0$, and a triplet subspace corresponding to $C_{12}=2$. 
Here, singlet means symmetric with respect to the 
exchange operator $P_{12}$ whereas triplet means antisymmetric.

We can now define precisely the basis set corresponding to the physical states 
with angular momentum $M_L$
and either singlet or triplet exchange symmetry. 
Since the quadruplet of indices 
$(n_1,n_2,n_3,n_4)$ and $(n_3,n_4,n_1,n_2)$ give the 
same symmetrised ket in equation (\ref{sym-ket}), 
we have only to consider one of 
the two quadruplet to label uniquely the symmetrised basis. Consequently, 
for singlet states, we set: 
\begin{eqnarray}
\mathcal{B}_{M_L}^{\mathrm{sym}}=
&&\left\{|n_1,n_2,n_3,n_4\rangle^+,\ n_1-n_2+n_3-n_4=4M_L,n_i \ge 0,\right.\nonumber\\
&&C_{12}=0,
         \left.(n_1>n_3 \ \mathrm{or}\ (n_1=n_3 \ \mathrm{and}\  n_2 \geq n_4)  \right\},\label{basis-sym}
\end{eqnarray}
and for triplet states:
\begin{eqnarray}
\mathcal{B}_{M_L}^{\mathrm{anti-sym}}=
&&\left\{|n_1,n_2,n_3,n_4\rangle^+,\ n_1-n_2+n_3-n_4=4M_L,n_i \ge 0,\right.\nonumber\\
&&C_{12}=2,
         \left.(n_1>n_3 \ \mathrm{or}\  (n_1=n_3 \ \mathrm{and}\  n_2>n_4)  \right\}.\label{basis-as}
\end{eqnarray}

\subsubsection{Selection rules and matrix elements}
\begin{figure}
\includegraphics[width=9cm]{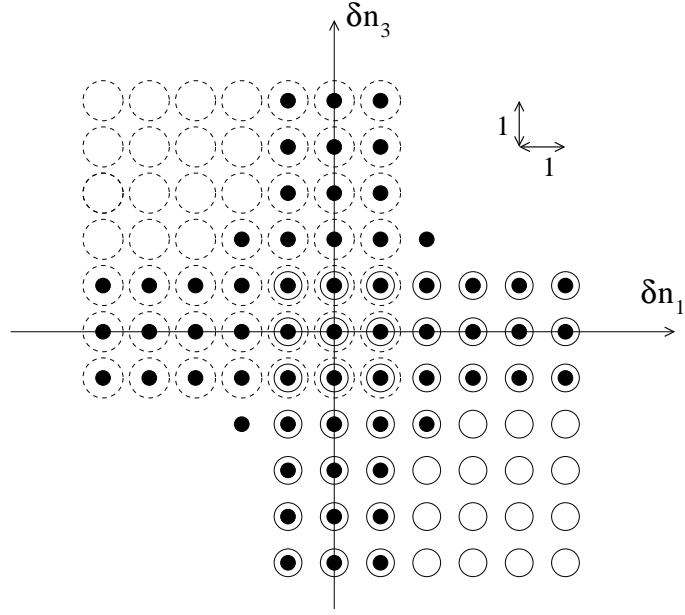}
\caption{\label{select-rules}The 159 selection rules 
that preserve the exchange symmetry are depicted in the $(n_1,n_3)$ space. The dark circles 
corresponds to the 61 rules $\delta n_1=\delta n_2$
and $\delta n_3=\delta n_4$, 
the full line circles to the 49 rules $\delta n_1-\delta n_2=4$ 
and $\delta n_3-\delta n_4=-4$ and the dashed line circles to
the 49 rules $\delta n_1-\delta n_2=-4$ and $\delta n_3-\delta n_4=4$.}
\end{figure}

Two basis vectors $|n_1,n_2,n_3,n_4\rangle$ and 
$|n_1+\delta n_1,n_2+\delta n_2,n_3+\delta n_3,n_4+\delta n_4\rangle$ 
are coupled by the Hamiltonian if the shifts $\delta n_i$ correspond to 
one of the 225 allowed coupling rules. Because the Hamiltonian commutes with the 
total angular momentum, they all obey $\delta n_1-\delta n_2+\delta n_3-\delta n_4=0$.
Among them, 159 rules preserve the exchange symetry while 
66 do not. The 159 rules that appear for the operators 
$T_1+T_2$, $T_{12}$, $(r_1+r_2)r_{12}$, $r_1r_2$ and $B$, 
obey $\delta n_1-\delta n_2=-(\delta n_3-\delta n_4)=0$ 
or $\delta n_1-\delta n_2=-(\delta n_3-\delta n_4)=\pm4$, and are shown in figure \ref{select-rules}.  
The 66 ones verify 
 $\delta n_1-\delta n_2=-(\delta n_3-\delta n_4)=\pm2$. 
They appear if the exchange symmetry is broken ($m_1 \neq m_2$ or $Q_1 \neq Q_2$) 
in the kinetic terms $T_1$, $T_2$, and the potential terms $r_1r_{12}$ and $r_2r_{12}$.
 
Since the Hamiltonian has been written in normal order, the derivation 
of the matrix elements is straigtforward. 
They are too numerous to be written explicitely 
here~\footnote{An alternative method
for computing matrix elements is to use the explicit expressions of the
basis states wavefunctions, and compute the action of the various
operators using well-known recursion relations between Laguerre polynomials.
The algebraic method that we use is much safer.}.
We only give two matrix 
elements of the kinetic operator of the 2D helium $T_1+T_2$ between 
two unsymmetrised basis vectors:
\begin{eqnarray}
&&\langle n_1,n_2,n_3,n_4|(T_1+T_2)|n_1,n_2,n_3,n_4\rangle=\\
&&1/4\,\left ({ n_1}+{ n_2}+1\right )\left ({ n_3}+{ n_4}+1
\right ) \nonumber\\
&&\ \ \left ({{ n_1}}^{2}+4\,{ n_1}\,{ n_2}-{ n_1}\,{ n_3}
+{ n_1}\,{ n_4}+{{ n_2}}^{2}+{ n_2}\,{ n_3}-{ n_2}\,{ 
n_4}\right.\nonumber\\
&&\left.+{{ n_3}}^{2}+4\,{ n_3}\,{ n_4}+{{ n_4}}^{2}+3\,{ n_1}+3
\,{ n_2}+3\,{ n_3}+3\,{ n_4}+8\right ),\nonumber
\end{eqnarray}
as well as the matrix element corresponding to the selection rule 
$\delta_1=\delta_2=\delta_3=\delta_4=1$:
\begin{eqnarray}
&&\langle n_1\!+\!1,n_2\!+\!1,n_3\!+\!1,n_4\!+\!1|(T_1+T_2)|n_1,n_2,n_3,n_4\rangle=\\
&&-1/4\,\sqrt {{ n_4}+1}\sqrt {{ n_3}+1}\sqrt {{ n_2}+1}\sqrt {{
 n_1}+1} \nonumber\\
&& \left ({{ n_1}}^{2}+5\,{ n_1}\,{ n_2}-2\,{ n_1}\,{
 n_4}+{{ n_2}}^{2}-2\,{ n_2}\,{ n_3}+{{ n_3}}^{2}+5\,{ n_3
}\,{ n_4}\right.\nonumber\\
&&\left.+{{ n_4}}^{2}+5\,{ n_1}+5\,{ n_2}+5\,{ n_3}+5\,{
 n_4}+12\right).\nonumber
\end{eqnarray}
\subsubsection{Numerical implementation}
For the numerical calculations, we have chosen to truncate the basis defined 
by equation (\ref{basis-sym}) or (\ref{basis-as}) using the condition $n_1+n_2+n_3+n_4 \leq N_{base}$. 
Because the angular momentum is fixed, we have only 3 independant indices,
and the size of the basis is roughly  $N_{base}^3/192$.
The basis $\mathcal{B}$ is then ordered in order to represent 
the Schr\"{o}dinger equation using band matrices as narrow as possible. 
The eigenvalue problem
is then solved using the Lanczos algorithm \cite{ericsson} which makes it possible
to compute a few eigenvalues in the range of interest.

\subsubsection{Variational parameter}
So far, the natural length scale of the problem is the Bohr radius $a_0$.
 Because it is not necessarilly the best suited one, 
 we introduce the length scale as $\alpha^{-\frac{1}{4}} a_0$. 
 The Schr\"{o}dinger equation 
(\ref{Schro-equ}) writes:
\begin{equation}
\left\{\alpha^{4}\left(\frac{T_1}{2\mu_{13}}+\frac{T_2}{2\mu_{23}}
+\frac{T_{12}}{m_3}\right)
+\alpha^{8}V\right\}|\Psi\rangle=\alpha^{12}\ E\ B\ |\Psi\rangle. \label{Schro-equ-alpha}
\end{equation}
When the basis is truncated, the length scale $\alpha$ becomes a 
variational parameter (i.e. the calculated energy levels should not
depend on $\alpha$ is the basis set is large enough) 
that has to be numerically optimized. 
All the numerical results presented in this paper are 
obtained with $\alpha$ close to 0.4. 
All the digits of the energy levels given in the Tables are significant.
The uncertainty on the results is thus 1 on the last figure,
and the relative accuracy reaches the $10^{-13}$ level.
\subsection{The 2D helium atom without electron interaction}
Let us consider the 2D helium with a fixed nucleus
of charge $Q_3=2$ (the mass $m_3$ is infinite). 
The Schr\"{o}dinger equation 
equation (\ref{Schro-equ-alpha}) simply writes :
\begin{equation}
\left\{\alpha^4\frac{T_1+T_2}{2}
+\alpha^8V\right\}|\Psi\rangle=\alpha^{12}\ E\ B\ |\Psi\rangle,
\end{equation}
where $V=-32\ (r_1+r_2)r_{12}+16 r_1r_2$.
If the $16 r_1 r_2$ term in the potential energy is removed, 
the three body problem corresponds to two independent 2D
hydrogen atoms with a nucleus of charge $Q=2$. The spectrum 
of the 2D hydrogen atom is well known, and is given by 
the series \cite{englefield}:
\begin{equation}
E_{N,M}=-\frac{Q^2}{2(N-1/2)^2}, \label{energyH2d}
\end{equation}
where $N \geq 1$ is the principal quantum number and 
$-N\!+\!1 \leq M \leq N\!-\!1$ 
the angular momentum of the electron; the degeneracy is $2N\!-\!1$. 
The structure of the energy spectrum is very similar to the 3D energy spectrum,
the only difference being that the effective quantum number $N-1/2$ is a half-integer
ranging from 1/2 to infinity rather than a non-negative integer.

For the helium atom without electronic interaction, 
the spectrum is thus given by:
\begin{equation}
E_{N_1,M_1}+E_{N_2,M_2}=-\frac{4}{2(N_1-1/2)^2}-\frac{4}{2(N_2-1/2)^2},
\end{equation}
where $N_1$ and $N_2$ are the principal quantum numbers 
of the two electrons. The essential degeneracy  is 
$2(2N_1\!-\!1)(2N_2\!-\!1)$ if $N_1\neq N_2$ and $(2N_1\!-\!1)^2$ otherwise
\footnote{Accidental degeneracies may occur for particular couples of principal quantum numbers.}.
The total angular momentum is simply given by $M_L=M_1+M_2$.
The states of total angular momentum $M_L$
correspond to the indices $(N_1,M_1,N_2,M_2)$ and $(N_2,M_2,N_1,M_1)$. 
Those degenerate states give symmetric (singulet) and 
antisymmetric (triplet) states when the two quadruplet are different
and only one symmetric state if they are equal. Finally, the energy 
levels can be labelled by $N_1,N_2,M_L$ and $P_{12}$. The degeneracy 
of this configuration is given by the number of solutions of $M_L=M_1+M_2$
taking into account the boundaries on $M_1$ and $M_2$.

By solving the Schr\"{o}dinger equation for an angular momentum
between -3 and 3, and for the two exchange symmetries, we have checked 
that our method 
gives the expected eigenenergies and degeneracies.

We have then checked the effect of the electronic interaction 
by introducing it perturbatively as $\epsilon/r_{12}$. We have 
numerically computed 
the ground state
energy of the three body problem as a function of $\epsilon$ and 
observed a linear behaviour,
as expected from first order 
perturbation theory. The slope in atomic units is $4.70(1)$, 
in agreement with the slope $3\pi/2$ predicted by  
first order perturbation theory (see appendix \ref{appendix-pente}).

\subsection{The 2D helium atom}
The $1/r_{12}$ term describing the electronic repulsion is now taken 
into account.
This does not affect the positions of the various ionization thresholds
(as the electron interaction vanishes at large distance). There is an infinite
number of single ionization thresholds associated with the
principal quantum number of the hydrogenic state of the resulting
$\mathrm{He}^{+}$ ion, given by energies:
\begin{equation}
I_N = - \frac{4}{2(N-1/2)^2}\ .
\end{equation}
These single ionization thresholds form a series which converge to the
double ionization threshold at zero energy.

Consequently, one expects bound states below energy $I_1=-8$ a.u.,
resonance (doubly excited states) between $I_1$ and zero, and only continua
above.
 
\subsubsection{Bound states}
The lowest energy levels of the 2D helium below the first ionization
limit are given in 
Table \ref{singulet} for the singlet states and in Table \ref{triplet}
for the triplet states. 
For each value of $M_L$, we obtain a Rydberg series converging
to the $N=1$ threshold.
For such excited states, the outer electron lies far from
the nucleus while the inner electron is essentially in its
ground state and lies very close to the nucleus. Because this picture gives two
very different roles to the two electrons, it results in a new
set of quantum numbers, namely $(N,M)$ for the inner
electron and $(n,m)$ for the outer one.
A brutal but
useful approximation is to neglect the effect
of the outer electron on the inner one, i.e. consider that
the inner electron in the hydrogenic state $N=1,M=0$ while the outer electron
sees a point charge $Q=1$ (the charge +2 of the nucleus screened by the charge
-1 of the inner electron)
at the origin, resulting in an energy spectrum $-8-1/(2(n-1/2)^2),$ where $n$
is the principal quantum number of the (hydrogenic) outer electron.
This is of course only an approximation. Deviations from it can be
measured through the quantum defect $\delta_{n,m}$ defined directly from
the energy levels through:
 \begin{equation}
E_{1,0,n,m}=-8-\frac{1}{2(n-1/2-\delta_{n,m})}\ .\label{Edelta}
\end{equation}
\begin{table}
\begin{tabular}{cclcc}
\hline
$N,M,n,m$ & $M_L$ & Energy (a.u.) & $\delta_{n,m}$\\
\hline
$1,0,1,0$ & 0 & -11.899 822 342 953& 0.1419\\
$1,0,2,0$ & 0 & -8.250 463 875 379 & 0.0871\\
$1,0,3,0$ & 0 & -8.085 842 792 777 & 0.0866\\
$1,0,4,0$ & 0 & -8.042 911 011 139 & 0.0865\\
$1,0,5,0$ & 0 & -8.025 668 309 76  & 0.0864\\
$1,0,6,0$ & 0 & -8.017 061 08      & 0.0864\\
\hline
$1,0,2,1$ & 1 & -8.211 542 089 886 & -0.0374\\
$1,0,3,1$ & 1 & -8.077 637 328 985 & -0.0378\\
$1,0,4,1$ & 1 & -8.039 947 878     & -0.0378\\
$1,0,5,1$ & 1 & -8.024 280 94      & -0.0379\\
\hline
$1,0,3,2$ & 2 & -8.079 805 619 119 & -0.0030\\
$1,0,4,2$ & 2 & -8.040 745 817 &     -0.0030\\ 
$1,0,5,2$ & 2 & -8.024 657 76&       -0.0031\\
$1,0,6,2$ & 2 & -8.016 51  &         -0.0031\\
$1,0,7,2$ & 2 & -8.011 80 &   -\\
\hline
\end{tabular}
\caption{\label{singulet}Energy levels of the singlet states of the 2D helium 
atom (with infinite mass of the nucleus),
below the first ionization threshold.
The optimum variational parameter $\alpha$ is close to
0.4. For most of the states, the basis truncation is given 
by $N_{base}=200$.
The basis size is then 43626 for singlet $M_L=0$ states, and slightly 
decreases with $M_L$. For the $(1,0,4,0), 
(1,0,5,0)$ and $(1,0,6,0)$ we use 
$N_{base}=240$ and a basis size of 74801. In the fourth column, 
$\delta_{n,m}$ is the quantum defect of the state, as deduced from 
equation~(\ref{Edelta}).}
\end{table}

\begin{table}
\begin{tabular}{cclc}
\hline
$N,M,n,m$ & $M_L$ & Energy (a.u.) & $\delta_{n,m}$ \\
\hline
$1,0,2,0$ & 0 & -8.295 963 728 090 & 0.2002\\
$1,0,3,0$ & 0 & -8.094 583 618 582 & 0.2008\\
$1,0,4,0$ & 0 & -8.045 941 305 572 & 0.2010\\
$1,0,5,0$ & 0 & -8.027 055 169     & 0.2011\\
$1,0,6,0$ & 0 & -8.017 807         & 0.2011\\
\hline
$1,0,2,1$ & 1 & -8.225 772 173 259 & 0.0118\\
$1,0,3,1$ & 1 & -8.080 919 691 737 & 0.0142\\
$1,0,4,1$ & 1 & -8.041 165 882 92  & 0.0149\\
$1,0,5,1$ & 1 & -8.024 858 500     & 0.0152\\
\hline
$1,0,3,2$ & 2 & -8.079 819 688 304 & -0.0028\\
$1,0,4,2$ & 2 & -8.040 751 693 48 &  -0.0028\\
$1,0,5,2$ & 2 & -8.024 661 158 &     -0.0028\\
$1,0,6,2$ & 2 & -8.016 512 &         -0.0028\\
\hline
\end{tabular}
\caption{\label{triplet}Energies of the triplet states
of the 2D helium atom (with infinite mass of the nucleus),
below the first ionization threshold.
The optimum variational parameter $\alpha$ is close to
0.4. The basis truncation is given by $N_{base}=200$.
The basis size is 43550 for triplet $M_L=0$ states.
In the fourth column, 
$\delta_{n,m}$ is the quantum defect of the state, as deduced from 
equation~(\ref{Edelta}).}
\end{table}

\begin{table}
\begin{tabular}{cc|cc}
\hline
\multicolumn{2}{c|}{3D Rydberg series}&\multicolumn{2}{c}{2D Rydberg series}\\
\hline
 & $\delta$ & $N,M_L$ & $\delta$ \\
$^1S^e$ & 0.140  & 1,0 & 0.0864 \\
$^3S^e$ & 0.299  & 1,0 & 0.2011 \\
$^1P^o$ & -0.012 & 1,1 & -0.0379\\
$^3P^o$ & 0.068  & 1,1 & 0.0152\\
$^1D^e$ & 0.0021 & 1,2 & -0.0031\\
$^3D^e$ & 0.0028 & 1,2 & -0.0028\\
\hline
\end{tabular}
\caption{\label{defauts-q}Quantum defects for various series 
of the 2D and 3D helium atoms below the first ionization threshold. 
The values
in the 3D case are calculated from the energies given 
in \cite{levin-micha}. The values in the 2D case are
the limits of $\delta_{n,m}$ for 
large values of $n$.}
\end{table}
If the previous approximation were exact, the quantum defects will all be zero.
Hence, deviations from zero and evolutions with $n$ and $m$ directly measure
the breaking of the approximation.
The results shown in Tables \ref{singulet} and \ref{triplet} show that
-- alike the 3D helium atom -- the quantum defect in a given series tend
to a constant value as $n\rightarrow \infty.$ 
When $|m|$ is increased, the outer electron is repelled
from the nucleus by the centrifugal energy barrier and fills less the
presence of the inner electron. It is thus expected that the quantum defects
will decrease with increasing $|m|$ and this is fully confirmed by our
``exact" diagonalizations, see Tables \ref{singulet}, \ref{triplet} and \ref{defauts-q}.
Also, in the triplet states, the wave function in configuration space is
antisymmetric, so that the two electrons cannot seat at the same place.
Consequently, they feel each other less efficiently, resulting in
a lower interaction energy and consequently a larger quantum defect. 
Again, our ``exact" calculations confirm this behaviour.

Finally, it is interesting to compare the 2D and 3D situations. Although
the Rydberg series are similar in both cases, this is not true for the ground state.
Indeed, the binding energy of the inner electron in its ground state is $8\ a.u.$, 
see equation (\ref{energyH2d}), in 2D,
that is four times more 
than in the 3D helium atom. Almost the same ratio 4 is observed between
the total binding energy of the ground state: 11.90 a.u. in 2D versus
2.91 a.u. in 3D.

For singly excited states, the core is also 4 times smaller in 2D.
We then expect a smaller core penetration because the centrifugal barrier 
is almost the same in the 2D and 3D systems, and also a smaller 
core polarisation
by the outer electron, resulting in smaller quantum defects in the 2D case.  
The comparison of
the 2D and the 3D quantum defects in Table \ref{defauts-q} is consistent
with this interpretation.

\subsubsection{Resonances}
Above the first ionization threshold, the spectrum contains resonances 
embedded into the continuum. They can be numerically separated using 
the complex rotation method \cite{complexrotation-Ho,complexrotation-Balslev} (also known
as the method of complex coordinates), where the positions 
and momenta 
$\bf r$ and $\bf p$ are reespectively changed into
${\bf r} e^{i\theta}$ and ${\bf p} e^{-i\theta}$. 
Here, it is simply implemented using a
complex length scale $\alpha=|\alpha|e^{i\theta}$ (in order to 
preserve their canonical commutation relations) This results in
a ``complex rotated" non hermitian Hamiltonian whose eigenvalues
are complex.
In the complex energy plane, the resonances do not depend on 
the angle $\theta$
while the continua are rotated by an angle of $2\theta$ around the
ionization thresholds.
The first resonance of the 2D helium atom (infinite mass of the nucleus) 
is obtained for zero angular momentum 
and singlet exchange symmetry. Its energy is:
\begin{equation}
E=-1.411\ 496\ 328 (1) - i \ 0.001\ 241\ 734 (1) \mathrm{a.u.}
\end{equation}
It is obtained for a rotation angle $\theta \approx 0.4$, 
a length scale $\alpha \approx 0.35$,
$N_{base}\!=\!150$, 
and a basis size of 18696.

\begin{figure}
\includegraphics[width=9cm]{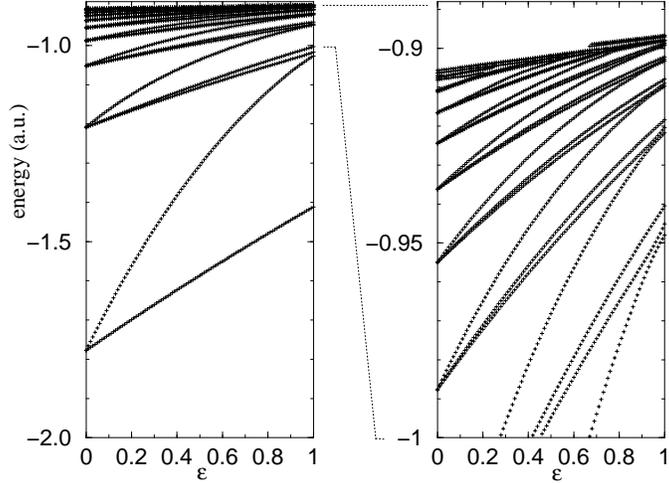}
\caption{\label{energies-epsilon}Energy levels of the 
singlet $M_L=0$ resonances of the  
2D helium atom (nucleus with infinite mass) between the first (-8 a.u.) 
and the second (-0.888$\dots$ a.u.)
ionisation thresholds as a function of the magnitude $\epsilon$ of the electronic 
interaction included in the potential energy as $\epsilon/r_{12}$.
The right part of the figure is a zoom close to the second ionisation threshold.
 All the 
energies presented on this picture are well converged and obtained with $\alpha=0.5,$ 
 $\theta=0.12$, $N_{base}=150$. The basis size is 18696. Because the method of complex
 coordinates is used, these energy levels are not bound states, but resonances.
 At the scale of this figure, their widths are very small.}
\end{figure}

The energy structure of the resonances is illustrated in the 
case of the singlet $M_L=0$ states in
figure \ref{energies-epsilon}. The electronic repulsion
is included in the potential energy as 
$\epsilon/r_{12}$ with $0\leq \epsilon \leq 1.$
One can follow the 
energy levels as a function of $\epsilon$ from the 
independent electron case 
($\epsilon\!=\!0$) to the helium case ($\epsilon\!=\!1$).
For $\epsilon=0$, the levels correspond to the Rydberg series 
$N=2$, $n\geq2$ converging to $I_2=-8/9$ a.u.. 
The degeneracy of the $(N=2,n=2)$ configuration is 9, 
with 3 states of zero total angular momentum. Two of them have 
the singlet symmetry and one the triplet symmetry. For $n>2$, 
the degeneracy of the configuration is 18, with 6 states corresponding 
to $M_L=0$ (3 singlet and 3 triplet states). Consequently, for $\epsilon=0$,
the first $M_L=0$ singlet resonance is doubly degenerate, 
 and the following ones are triply degenerate. 
The introduction of the electronic interaction removes
the degeneracy. 

\subsection{The 2D $H^-$ ion}
The $H^-$ ion with a fixed nucleus is obtained by setting $Q_3=1$.
We obtain only one bound state below the first ionisation limit 
(at -2 atomic units from equation (\ref{energyH2d})), 
with zero angular momentum 
and singlet exchange symmetry. Its energy is:
\begin{equation}
E\ =\ -2.240\ 275\ 363\ 589 (1) \mathrm{a.u.}
\end{equation}
It is obtained for $\alpha \approx 0.4$, $N_{base}=220$,
the basis size being 57820. 

\section{Conclusion}
We have introduced a new set of coordinates 
to represent the 2D three body Coulomb problem and given the 
resulting Schr\"{o}dinger equation.
We have discussed the discrete symmetry group properties 
of the equation and shown that it can be numerically solved 
very efficiently, using a convenient basis set for which the 
Schr\"{o}dinger equation involves sparse banded matrices.
The convergence of the calculations is very good and the 
numerical results are extremely accurate. This is demonstrated
in the case of the 2D Helium atom (with inifinite mass
of the nucleus), for which
the lowest energy levels in the bound Rydberg series 
are given
with a relative accuracy in the $10^{-9}$ to $10^{-13}$ range.

The method developped in this paper provides 
an efficient tool for studying the dynamics of the 2D Helium atom 
in an external electric field aligned along the $x$ axis. 
Indeed, with a field of strength $F$, we must add the external 
potential energy term $V_{ext}= 16\ r_1r_2r_{12} (x_1+x_2)\ F$ 
to equation (\ref{potentiel}). 
The only remaining symmetries are then the 
 exchange symmetry $P_{12}$ and the symmetry with respect 
 to the $x$ cartesian axis $\Pi_x$. In such a case, the convenient basis 
 set can be defined, from equation (\ref{sym-ket}), by:
 \begin{equation}
|n_1,n_2,n_3,n_4\rangle^{\epsilon}=|n_1,n_2,n_3,n_4\rangle^{+}
                         +\epsilon \ |n_2,n_1,n_4,n_3\rangle^{+},
\end{equation}
with $\epsilon=\pm 1$ for even or odd states with respect to $\Pi_x$.
Because $V_{ext}$ is a polynomial in the new coordinates, it exhibits
 selection rules, making an accurate diagonalisation of 
 the Schr\"{o}dinger equation still possible.

The motion of the nucleus can easily be taken into account, 
including the $T_{12}$ contribution to the Hamiltonian 
in equation (\ref{Schro-equ}). That way, it will be possible to determine 
very accurately the ground state energy of excitonic trions, 
as a function 
of the electron to hole mass ratio. This work is in progress.

\begin{acknowledgments}
Laboratoire Kastler Brossel de
l'Universit\'e Pierre
et Marie Curie et de l'Ecole Normale Sup\'erieure is
UMR 8552 du CNRS. 
CPU time on various computers has been provided by IDRIS.
The authors acknowledge Monique Combescot for stimulating 
discussions on the trion problem.
\end{acknowledgments}

\appendix
\section{Character table\label{table-xi}}
The discrete symmetry group G of the Schr\"{o}dinger equation (\ref{Schro-equ}) 
written in the $(x_p,y_p,x_m,y_m)$ coordinates  is studied here.
It is generated by the $\Pi_x$, $\Pi_y$, $P_{12}$ physical symmetries and 
the $\Pi_1$, $\Pi_2$, $\Pi_p$, $\Pi_m$ ``additional" symmetries, defined in equations 
(\ref{pixpiy}), (\ref{P12}) and (\ref{pi1pi2pippim}).
Its structure has been studied following standard methods of the theory of finite groups 
(see for example \cite{petrachene,hamermesh}).
Because all the generators of the group
can be seen as a permutation among the sixteen quantitites
$x_p,y_p,x_m,y_m,(x_p+y_p)/\sqrt{2},(x_p-y_p)/\sqrt{2},(x_m+y_m)/\sqrt{2},(x_m-y_m)/\sqrt{2}$ 
and the opposite values, 
the G group appears as a subgroup  
of the permutation group of sixteen elements.
Thanks to this property, it has been easily studied using the permutation group package provided 
by the {\it Maple} language.

The group G contains 128 elements, in 29 classes. It has 16 one-dimensional, 
8 two-dimensional and 5 four-dimensional irreductible representations.
Table \ref{table-car} represents its complete character table.
It has been obtained using the method described in \cite{hamermesh}. Let 
the classes of G be
$K_i$ with $1 \leq i \leq n$. The set $K_iK_j$, set of the products of 
any element of the class $K_i$ by any element
of the class $K_j$, is made of complete classes. Calling $c_{ijl}$ the 
number of occurrences
of the class $K_l$ in the product $K_iK_j$, one can symbolically write:
\begin{equation}
K_iK_j=\sum_{l}c_{ijl}K_l.
\end{equation}
This property is used to obtain relations between characters $\chi_i^{(R)}$ 
of the class $K_i$
in an irreductible representation $R$ :
\begin{equation}
g_ig_j\ \chi_i^{(R)}\chi_j^{(R)}=\chi_E^{(R)}\sum_{l=1}^{29}c_{ijl}\ g_l\chi_l^{(R)}.\label{system}
\end{equation}
where , $g_i$ is the number of elements of the class $K_i$ and 
$\chi_E^{(R)}$ is the character of the 
identity E, {\it i.e.} the dimension of the representation $R$.
The $n$ characters of an irreductible representation appear to be
 the solutions of the
$n(n+1)/2$ quadratic equations obtained from (\ref{system}) for any couple of $(i,j)$.
Then, to construct the character table of the group,
three steps are necessary:
first, the group has to be separated in classes, and the number $g_i$ are obtained. 
Second, the numbers $c_{ijl}$ are computed, 
and last the system of equations \ref{system} is solved.
Obviously, there $n$ different sets of solutions, 
corresponding to the $n$ irreductible representations.
\section{\label{appendix-pente}}
The scalar product of two wave functions $|\Psi^{(1)}\rangle$ 
and $|\Psi^{(2)}\rangle$ in the 
$(x_p,y_p,x_m,y_m)$ coordinates is :
\begin{equation}
\langle\Psi^{(1)}|\Psi^{(2)}\rangle=\frac{1}{16}\int\!\!\!\int\!\!\!\int\!\!\!\int\Psi^{(1)}(x_p,y_p,x_m,y_m)^*\ B\ \Psi^{(2)}(x_p,y_p,x_m,y_m)\ 
\ dx_p\ dy_p\ dx_m\ dy_m,
\end{equation}
where $B$ is given by equation (\ref{Bmat}).
The integrals are calculated from 
$-\infty$ to $+\infty$. The factor $1/16=1/2^4$ comes from the 
four double mappings of the space introduced by the change to
parabolic coordinates.

We now calculate the average value of $1/r_{12}$
for the ground state $|\Psi_{0,0}\rangle$ of a 2D helium atom 
without electronic repulsion, that is:
\begin{equation}
\sigma=\left\langle\Psi_{0,0}\left|\frac{1}{r_{12}}\right|\Psi_{0,0}\right\rangle.
\end{equation} 
The normalised wave function of the ground state 
of a 2D hydrogenic atom with a
 nucleus of charge Q is :
\begin{equation}
\Psi_0(r_1)=\sqrt{\frac{2}{\pi}}\ 2Q\ e^{-2Q r_1},
\end{equation}
so that:
\begin{equation}
\Psi_{0,0}(r_1,r_2)=\Psi_{0,0}(r_1)\ \Psi_{0,0}(r_2)
                     =\frac{8Q^2}{\pi}\ e^{-2Q (r_1+r_2)}. \label{psi00}
\end{equation}
We now evaluate $\sigma$ using the 
$(x_p,y_p,x_m,y_m)$ coordinates.
Since the jacobian of the coordinate transformation is $B=16r_1r_2r_{12}$,
$\sigma$ writes:
\begin{equation}
\sigma=\int{\!\!\!\int{\!\!\!\int{\!\!\!\int{16 r_1r_2 
|\Psi_{0,0}(x_p,y_p,x_m,y_m)|^2\ dx_p\ dy_p\ dx_m\ dy_m}}}}\label{sigma},
\end{equation} 
where $r_1$ and $r_2$ are given by equations (\ref{distances}). 
The integrals are calculated from $-\infty$ to $+\infty$.

To evaluate $\sigma$, we represent the $(x_p,y_p)$ and $(x_m,y_m)$ planes using polar 
coordinates $(r_p,\theta_p)$ and $(r_m,\theta_m)$ and obtain from equation (\ref{distances}):
\begin{eqnarray}
r_1&=&\frac{1}{16}\left(r_p^4+r_m^4+2r_p^2r_m^2cos(2\theta_p-2\theta_m) \right)\nonumber\\
r_2&=&\frac{1}{16}\left(r_p^4+r_m^4-2r_p^2r_m^2cos(2\theta_p-2\theta_m) \right)\\
r_{12}&=&\frac{r_p^2r_m^2}{4}\nonumber
\end{eqnarray}
The ground state wave function is then:
\begin{equation}
\Psi_{0,0}(r_p,\theta_p,r_m,\theta_m)=\frac{8Q^2}{\pi}\ e^{-Q(r_p^4+r_m^4)/2},
\end{equation}
so that :
\begin{equation}
\sigma=\frac{Q^4}{4\pi^2}\int{\!\!\!\int{\!\!\!\int{\!\!\!\int{
\left(r_p^8+r_m^8+2r_p^4r_m^4\left(1-2cos^2(2\theta_p-2\theta_m)\right) \right)
e^{-Q(r_p^4+r_m^4)/2} 
\ r_pdr_p\ r_mdr_m\ d\theta_p\ d\theta_m}}}}.
\end{equation}
The integration over $\theta_p$ and $\theta_m$ gives 0 for the angular dependant term and $4\pi^2$ 
for the independant one.
The integration over $r_p$ and $r_m$ involves Gaussian integrals that give
\begin{equation}
\sigma=\frac{3\pi Q}{4},
\end{equation}
that is $3\pi /2$ when $Q=2$.
In the 3D case, $\sigma$ is evaluated to $5Q/8$ 
in \cite{betheetsalpeter}.
The ratio $\sigma_{2D}/\sigma_{3D}=6\pi/5\simeq 3.77$ is close to 4,
because the 2D ground state wave function is four times smaller 
than the 3D wave function. 



\begin{sidewaystable}
\tiny
\begin{ruledtabular}
\begin{tabular}{ccccccccrrrrrrrrrrrrrrrrrrrrr}
1&8&8&2&4&8&8&4&1&4&8&2&8&2&4&4&4&8&2&4&4&8&4&2&4&2&4&4&2\\
\hline
E&$\Pi_y$&$\Pi_x$&$\Pi_y\Pi_x$&$P_{12}$&$\Pi_yP_{12}$&$\Pi_xP_{12}$&$\Pi_y\Pi_xP_{12}$&&&&&&&&&&&&&&&&&&&$\Pi_2$&$\Pi_1$&$\Pi_p$\\
&&&&&&&&&&&&&&&&&&&&&&&&&&&&$\Pi_m$\\
\hline
{\bf 1} &{\bf 1} &{\bf 1} &{\bf 1} &{\bf 1} &{\bf 1} &{\bf 1} &{\bf 1} & 1 & 1 & 1 & 1 & 1 & 1 & 1 & 1 & 1 & 1 & 1 & 1 & 1 & 1 & 1 & 1 & 1 & 1 & 1 & 1 & 1 \\ 
{\bf 1} &{\bf -1} &{\bf 1} &{\bf -1} &{\bf 1} &{\bf -1} &{\bf 1} &{\bf -1} & 1 & 1 & 1 & 1 & 1 & -1 & 1 & 1 & -1 & -1 & 1 & -1 & -1 & -1 & -1 & -1 & -1 & -1 & 1 & 1 & 1 \\ 
{\bf 1} &{\bf 1} &{\bf -1} &{\bf -1} &{\bf 1} &{\bf 1} &{\bf -1} &{\bf -1} & 1 & -1 & 1 & 1 & -1 & -1 & 1 & -1 & 1 & -1 & 1 & 1 & -1 & 1 & -1 & -1 & -1 & -1 & 1 & 1 & 1 \\ 
{\bf 1} &{\bf -1} &{\bf -1} &{\bf 1} &{\bf 1} &{\bf -1} &{\bf -1} &{\bf 1} & 1 & -1 & 1 & 1 & -1 & 1 & 1 & -1 & -1 & 1 & 1 & -1 & 1 & -1 & 1 & 1 & 1 & 1 & 1 & 1 & 1 \\ 
{\bf 1} &{\bf 1} &{\bf 1} &{\bf 1} &{\bf -1} &{\bf -1} &{\bf -1} &{\bf -1} & 1 & 1 & -1 & 1 & -1 & -1 & -1 & 1 & 1 & 1 & 1 & 1 & -1 & -1 & 1 & 1 & -1 & -1 & 1 & 1 & 1 \\ 
{\bf 1} &{\bf -1} &{\bf 1} &{\bf -1} &{\bf -1} &{\bf 1} &{\bf -1} &{\bf 1} & 1 & 1 & -1 & 1 & -1 & 1 & -1 & 1 & -1 & -1 & 1 & -1 & 1 & 1 & -1 & -1 & 1 & 1 & 1 & 1 & 1 \\ 
{\bf 1} &{\bf 1} &{\bf -1} &{\bf -1} &{\bf -1} &{\bf -1} &{\bf 1} &{\bf 1} & 1 & -1 & -1 & 1 & 1 & 1 & -1 & -1 & 1 & -1 & 1 & 1 & 1 & -1 & -1 & -1 & 1 & 1 & 1 & 1 & 1 \\ 
{\bf 1} &{\bf -1} &{\bf -1} &{\bf 1} &{\bf -1} &{\bf 1} &{\bf 1} &{\bf -1} & 1 & -1 & -1 & 1 & 1 & -1 & -1 & -1 & -1 & 1 & 1 & -1 & -1 & 1 & 1 & 1 & -1 & -1 & 1 & 1 & 1 \\ \\
1 & 1 & 1 & 1 & 1 & 1 & 1 & 1 & 1 & -1 & -1 & 1 & -1 & 1 & 1 & -1 & -1 & -1 & 1 & -1 & -1 & -1 & 1 & 1 & -1 & 1 & -1 & -1 & 1 \\ 
1 & -1 & 1 & -1 & 1 & -1 & 1 & -1 & 1 & -1 & -1 & 1 & -1 & -1 & 1 & -1 & 1 & 1 & 1 & 1 & 1 & 1 & -1 & -1 & 1 & -1 & -1 & -1 & 1 \\ 
1 & 1 & -1 & -1 & 1 & 1 & -1 & -1 & 1 & 1 & -1 & 1 & 1 & -1 & 1 & 1 & -1 & 1 & 1 & -1 & 1 & -1 & -1 & -1 & 1 & -1 & -1 & -1 & 1 \\ 
1 & -1 & -1 & 1 & 1 & -1 & -1 & 1 & 1 & 1 & -1 & 1 & 1 & 1 & 1 & 1 & 1 & -1 & 1 & 1 & -1 & 1 & 1 & 1 & -1 & 1 & -1 & -1 & 1 \\ 
1 & 1 & 1 & 1 & -1 & -1 & -1 & -1 & 1 & -1 & 1 & 1 & 1 & -1 & -1 & -1 & -1 & -1 & 1 & -1 & 1 & 1 & 1 & 1 & 1 & -1 & -1 & -1 & 1 \\ 
1 & -1 & 1 & -1 & -1 & 1 & -1 & 1 & 1 & -1 & 1 & 1 & 1 & 1 & -1 & -1 & 1 & 1 & 1 & 1 & -1 & -1 & -1 & -1 & -1 & 1 & -1 & -1 & 1 \\ 
1 & 1 & -1 & -1 & -1 & -1 & 1 & 1 & 1 & 1 & 1 & 1 & -1 & 1 & -1 & 1 & -1 & 1 & 1 & -1 & -1 & 1 & -1 & -1 & -1 & 1 & -1 & -1 & 1 \\ 
1 & -1 & -1 & 1 & -1 & 1 & 1 & -1 & 1 & 1 & 1 & 1 & -1 & -1 & -1 & 1 & 1 & -1 & 1 & 1 & 1 & -1 & 1 & 1 & 1 & -1 & -1 & -1 & 1 \\ 
2 & 0 & 0 & -2 & 0 & 0 & 0 & 0 & 2 & 2 & 0 & 2 & 0 & 0 & 0 & -2 & 2 & 0 & -2 & -2 & 0 & 0 & 2 & -2 & 0 & 0 & 0 & 0 & -2 \\ 
2 & 0 & 0 & 2 & 0 & 0 & 0 & 0 & 2 & 2 & 0 & 2 & 0 & 0 & 0 & -2 & -2 & 0 & -2 & 2 & 0 & 0 & -2 & 2 & 0 & 0 & 0 & 0 & -2 \\ 
2 & 0 & 0 & 2 & 0 & 0 & 0 & 0 & 2 & -2 & 0 & 2 & 0 & 0 & 0 & 2 & 2 & 0 & -2 & -2 & 0 & 0 & -2 & 2 & 0 & 0 & 0 & 0 & -2 \\ 
2 & 0 & 0 & -2 & 0 & 0 & 0 & 0 & 2 & -2 & 0 & 2 & 0 & 0 & 0 & 2 & -2 & 0 & -2 & 2 & 0 & 0 & 2 & -2 & 0 & 0 & 0 & 0 & -2 \\ 
2 & 0 & 0 & 0 & 0 & 0 & 0 & -2 & 2 & 0 & 0 & -2 & 0 & 2 & 0 & 0 & 0 & 0 & 2 & 0 & -2 & 0 & 0 & 0 & 2 & 2 & 2 & -2 & -2 \\ 
2 & 0 & 0 & 0 & 0 & 0 & 0 & -2 & 2 & 0 & 0 & -2 & 0 & 2 & 0 & 0 & 0 & 0 & 2 & 0 & 2 & 0 & 0 & 0 & -2 & 2 & -2 & 2 & -2 \\ 
2 & 0 & 0 & 0 & 0 & 0 & 0 & 2 & 2 & 0 & 0 & -2 & 0 & -2 & 0 & 0 & 0 & 0 & 2 & 0 & 2 & 0 & 0 & 0 & -2 & -2 & 2 & -2 & -2 \\ 
2 & 0 & 0 & 0 & 0 & 0 & 0 & 2 & 2 & 0 & 0 & -2 & 0 & -2 & 0 & 0 & 0 & 0 & 2 & 0 & -2 & 0 & 0 & 0 & 2 & -2 & -2 & 2 & -2 \\ 
4 & 0 & 0 & 0 & 0 & 0 & 0 & 0 & 4 & 0 & 0 & -4 & 0 & 0 & 0 & 0 & 0 & 0 & -4 & 0 & 0 & 0 & 0 & 0 & 0 & 0 & 0 & 0 & 4 \\ 
4 & 0 & 0 & $-\sqrt{8}$ & 2 & 0 & 0 & 0 & -4 & 0 & 0 & 0 & 0 & $\sqrt{8}$ & -2 & 0 & 0 & 0 & 0 & 0 & 0 & 0 & 0 & $\sqrt{8}$ & 0 & $-\sqrt{8}$ & 0 & 0 & 0 \\ 
4 & 0 & 0 & $\sqrt{8}$ & -2 & 0 & 0 & 0 & -4 & 0 & 0 & 0 & 0 & $\sqrt{8}$ & 2 & 0 & 0 & 0 & 0 & 0 & 0 & 0 & 0 & $-\sqrt{8}$ & 0 & $-\sqrt{8}$ & 0 & 0 & 0 \\ 
4 & 0 & 0 & $\sqrt{8}$ & 2 & 0 & 0 & 0 & -4 & 0 & 0 & 0 & 0 & $-\sqrt{8}$ & -2 & 0 & 0 & 0 & 0 & 0 & 0 & 0 & 0 & $-\sqrt{8}$ & 0 & $\sqrt{8}$ & 0 & 0 & 0 \\ 
4 & 0 & 0 & $-\sqrt{8}$ & -2 & 0 & 0 & 0 & -4 & 0 & 0 & 0 & 0 & $-\sqrt{8}$ & 2 & 0 & 0 & 0 & 0 & 0 & 0 & 0 & 0 & $\sqrt{8}$ & 0 & $\sqrt{8}$ & 0 & 0 & 0 \\ 
\end{tabular}
\end{ruledtabular}
\caption{\label{table-car}Character table of the discrete symmetry group G of the
Schr\"odinger equation (\protect\ref{Schro-equ}).
The classes and the irreductible 
representations have been organized in order to obtain the character 
table of the $D_{2h}$ group in the upper left corner (in bold figures). The last three 
classes are those of the ``additional" symmetries $\Pi_2$, $\Pi_1$ and $\Pi_p$ 
($\Pi_p$ and $\Pi_m$ belong to the same class). 
The first line gives the number of elements in each class.
We finally give one element of each of the 29 classes :
E, $\Pi_y$, $\Pi_x$, $\Pi_y\Pi_x$, $P_{12}$, $\Pi_yP_{12}$, 
$\Pi_xP_{12}$, $\Pi_y\Pi_xP_{12}$, $\Pi_p\Pi_m$, $\Pi_2\Pi_x\Pi_m$, $\Pi_2P_{12}$,
$\Pi_1\Pi_2$, $\Pi_2\Pi_xP_{12}$, $\Pi_p\Pi_y\Pi_xP_{12}$, $\Pi_pP_{12}$, $\Pi_2\Pi_x$,
$\Pi_2\Pi_y$, $\Pi_2\Pi_y\Pi_x$, $\Pi_1\Pi_2\Pi_m$, $\Pi_2\Pi_y\Pi_m$,
$\Pi_2\Pi_y\Pi_xP_{12}$, $\Pi_2\Pi_yP_{12}$, $\Pi_y\Pi_x\Pi_m$, $\Pi_p\Pi_y\Pi_x\Pi_m$,
$\Pi_1\Pi_y\Pi_xP_{12}$, $\Pi_y\Pi_xP_{12}^{-1}$, $\Pi_2$, $\Pi_1$, $\Pi_p$, 
some of them are reported in the second line of the Table.}
\end{sidewaystable}
\end{document}